\documentclass[12pt]{article}
\usepackage{fullpage,graphicx}
\usepackage{wrapfig}
\usepackage{amsmath}
\usepackage{cancel}
\usepackage{enumitem}
\usepackage{amssymb}
\usepackage{tabto}
\usepackage{cancel}
\usepackage{multirow}
\usepackage{booktabs}
\usepackage{makecell}
\usepackage{enumitem}
\usepackage{pdflscape}
\usepackage{tablefootnote}
\usepackage{xcolor}

\RequirePackage{caption}
\usepackage{cmap}
\usepackage[T1]{fontenc}
\begin{document}

{\large \bf The limited effect of electric conductivity on the ion current evaporated from electrospray sources}
\title{}

Ximo Gallud (ximogc@mit.edu)
and Paulo C. Lozano (plozano@mit.edu)
\\

Massachusetts Institute of Technology
\\
77 Massachusetts Ave, 02139, Cambridge, MA

\vspace*{1.0em}
\textbf{Abstract: Electrohydrodynamic modeling and experimental observations indicate that the ion evaporation current emitted by passively-fed electrospray sources is independent of the electrical conductivity of the ionic liquid.  This contrasts with cone-jet electrosprays, in which current depends on conductivity at fixed flow rates. The current in the pure ionic regime is controlled by the bias voltage and flow impedance, where a low viscosity is key for its enhancement. In addition to clarifying the role of electric conductivity in ionic liquid ion sources, this observation provides guidance to the selection of liquid properties for a particular application. 
}
\section{Introduction}
Room-temperature molten salts or ionic liquids are very useful in electrospraying for its enhanced ability to trigger ion emission from their surface through a process of field evaporation \cite{Iribarne1976}. Usually, the electrosprayed beam composition of these salts is a mixture of small droplets and evaporated ions, but under certain geometrical aspects of the emitters and physical conditions of the salts, it is possible to extinguish completely the droplet current at low enough flow rates $Q$ \cite{Romero-Sanz2003SourceRegime}.


The conditions that yield to a high current pure-ion beams in ionic liquids are not well understood and often rely on experimental heuristics, namely a high surface tension $\gamma$, and a high electrical conductivity $K$ \cite{Romero-Sanz2005IonicLiquids,Garoz2007TaylorConductivity,Lozano2006EnergySource,Castro2007EffectSourcesb,Larriba2007MonoenergeticLiquids,Castro2006CapillarySources}. The latter has specially served as a prescription recipe when looking for better ionic liquids: on top of the list are those that have the highest electrical conductivity when the objective is to maximize the emitted current, which is highly desirable for many applications. For instance, high current regimes are very appealing for high thrust space propulsion \cite{Chiu2007IonicPropulsion} or highly efficient microfabrication \cite{Takeuchi2013DevelopmentModification}, and ion etching \cite{Guilet2011IonicMachining,Yoshida2014ConcurrentArray}.

Although it is possible to reach the low $Q$ required for pure-ion electrospraying using a flow controller, engineering devices typically use passive flow emitters, which are naturally simpler, do not have moving parts, and are easily miniaturized to the scale where such low $Q$ is attainable. These passive flow sources usually consist of micrometer-sized capillaries or tips where the liquid feeding the electrospray meniscus flows across by either passing through the pores of these sources \cite{Perez-Martinez2015IonMicrotips} or forming liquid films that wet their surfaces \cite{Castro2009EffectTips}. Under these conditions, $Q$ is not predetermined by a controller, but instead results from a balance of electric, surface tension, and natural suction stresses induced by the hydraulic impedance of the capillaries or tips. 

From mass conservation arguments, the current emitted of electrosprays is proportional to $Q$, to the specific charge of the ions $\frac{q}{m}$, and density of the ionic liquid $\rho$ as:

\begin{equation}
    I = \rho Q \frac{q}{m}
    \label{eq:current_mass}
\end{equation}
The key question is, under these passive flow conditions, what is the role of the electric conductivity $K$ in the magnitude of the current?

If we assume the incompressibility of the liquid, any dependency of the current with the conductivity should act via a modification to $Q$, or the charge to mass ratio $\frac{q}{m}$.



 How much does the conductivity affect $Q$? Interestingly, in the cone-jet mode, there is clear experimental evidence that when $Q$ is not actively controlled, it does not depend on the conductivity of the liquid \cite{Smith2006VoltageElectrospraying,Smith2006TheIN, Ryan2014TheModes}. In fact, what confers the cone-jet current its characteristic square root dependence on the conductivity is the increase of specific charge $q/m$ via the reduction of the electrical relaxation time, and the reduction of the droplet radius. This can be seen if considering Gañán-Calvo's \cite{Ganan-Calvo1999TheLawsb} universal scaling law for the average surface charge $q_s$ and the diameter $d$ of the emitted droplets:
 \begin{equation}
 q_s = 0.59 \left(\varepsilon_0 \gamma^2 \rho K^2\right)^{\frac{1}{6}}
 \end{equation}
  \begin{equation}
d = 2.9 \; d_0 \; \sqrt{\frac{Q}{Q_0}} 
\label{eq:jet_diameter}
 \end{equation}
 Where $d_0 = \left(\frac{\varepsilon_0^2 \gamma}{\pi^2 \rho K^2}\right)^{\frac{1}{3}}$ is a characteristic diameter, $Q_0 = \frac{\varepsilon_0 \gamma}{\rho K}$ is a characteristic flow rate, $\gamma$ is the surface tension of the liquid and $\varepsilon_0$ is the dielectric permittivity of vacuum. In this case,

 \begin{equation}
     \frac{q}{m} = \frac{q_s S}{\rho V} = \frac{q_s \pi d^2}{\rho \pi d^3/6} = \frac{2.62}{\rho}\sqrt{\frac{\gamma K}{Q}}
     \label{eq:q_mGanan}
 \end{equation}

 Where $S$ and $V$ are the surface and volume of the droplet, assumed here as spherical. Inserting equation \ref{eq:q_mGanan} by $\rho Q$ in equation \ref{eq:current_mass} yields the well established scaling for the current \cite{FernandezdelaMora1994TheCones,Ganan-Calvo1997CurrentLaws}:

 \begin{equation}
 I \propto \sqrt{\gamma K Q}
 \label{eq:current_scaling}
 \end{equation}

Empirically, it is observed that  $\frac{q}{m}$ (or $K/Q$ from equation \ref{eq:q_mGanan}) cannot be indefinitely increased (for instance, by reducing $Q$ at constant $K$). The non-dimensional flow rate parameter,
\begin{equation}
\eta = \sqrt{\frac{\rho K Q}{\gamma \varepsilon \varepsilon_0}},
\label{eq:eta}
\end{equation}

serves as an empirical yardstick to predict the minimum flow rate at which cone-jet emission is possible \cite{FernandezdelaMora1994TheCones}. In general, such condition occurs when $\eta\sim 1$. However, for some liquids such as ionic liquids, it is possible to have large values of $K/Q$ while operating near $\eta \approx 1$, as the normal electric field becomes similar in magnitude to the critical field for ion evaporation \cite{Iribarne1976} $E^*$,
\begin{equation}
    E^* = \frac{4\pi\varepsilon_0 \Delta G^2}{q^3}
\end{equation}

Where $\Delta G$ is the energy of solvation of the ions (approximately 1-2 eV), and $q$ is the charge of the ion. If the flow rate is further decreased, the jet quenches up until the extent where all the droplet current is extinguished and the pure-ion regime is achieved, where all the flow rate is composed of different ion species that are evaporated from the apex of a closed surface.

Essentially, in the pure-ion mode, the specific charge  $\frac{q}{m}$ is the highest that an electrospray can have: that corresponding to the discrete nature of the ionic molecules. This magnitude of $\frac{q}{m}$ is obviously independent of the electric conductivity of the liquid. In the pure-ion regime, then, any dependence on the conductivity of the current emitted in passive flow ion sources must come from changes in the flow rate $Q$, given eq. \ref{eq:current_mass}.

\section{Scalings for $I$ and $Q$ in the pure-ion regime}
Scaling arguments have also emerged for the current and flow rate emitted in the pure ion regime \cite{Higuera,Coffman2019ElectrohydrodynamicsField}. For instance, it appears reasonable to assume that the amount of current emitted by an ionic liquid is established by the conductive transport to the surface in a region with a characteristic emission size $r^*$. 

This characteristic region of assumed spherical shape, could be approximated by a balance of electric and surface tension stresses at the emission interface,

\begin{equation}
\frac{1}{2}\varepsilon_0E^{*^2}\approx \frac{2\gamma}{r^*}.
\end{equation}
Where $E^*$ is the critical field for evaporation \cite{Iribarne1976}.
Solving for $r^*$,

\begin{equation}
r^*\;\approx\; \frac{4\gamma}{\varepsilon_0E^{*^2}}
\label{eq:rstar_2}
\end{equation}

Given that interfacial charges are not fully relaxed at this scale, an internal field that scales with $E^*/\varepsilon$ appears, thus driving a current density,

\begin{equation}
j = KE_{in}\approx K\frac{E^*}{\varepsilon}.
\label{eq:current_density}
\end{equation}

The current is then determined by the area of the emission region, which is also assumed to scale with the spherical cap area $2 \pi r^{*^2}$. 


\begin{equation}
I\;\approx\; jA \; \approx \; \frac{32\pi K\gamma^2}{\varepsilon_0^2E^{*^3}}\frac{\varepsilon}{(\varepsilon-1)^2}.
\label{eq:current_2}
\end{equation}

Since $E^*$ does not depend on $K$, this expression suggests that there is a direct link between the liquid conductivity and the emitted current of the ionic liquid. Analogously for the flow rate Q, we can apply eq. \ref{eq:current_scaling} to \ref{eq:current_2}:

\begin{equation}
Q \;\approx\; \frac{I}{\rho \frac{q}{m}} \; \approx \; \frac{32\pi K\gamma^2}{\rho \frac{q}{m}\varepsilon_0^2E^{*^3}}\frac{\varepsilon}{(\varepsilon-1)^2}.
\label{eq:flowrate_2}
\end{equation}

Scaling arguments like this, together with the interpretation of some empirical evidence, have fueled the notion that conductivity has a central role in the operation and performance of electrosprays in the pure ionic regime, more precisely through its effects on the magnitude of the emission current.

\section{The independence of $I$ and $Q$ on conductivity in the pure ion regime: electrohydrodynamic simulations}

Recent results from a detailed electrohydrodynamic numerical model  \cite{Gallud2022TheEvaporation} suggest that the scalings in eqs \ref{eq:current_2} and \ref{eq:flowrate_2} are somewhat incomplete, making them misleading at best.

The equilibrium shapes of ionic liquid menisci were computed in a parameter space of varying conductivities $K \in [0.9,2.4]$ S/m and viscosities $\mu \in [0.025,0.057]$ Pa$\cdot$s. The rest of physical properties  (surface tension, energy transport coefficients, mass density, emitted $q/m$, dielectric permittivity $\varepsilon$, $\Delta G$) were fixed to those similar to the ionic liquid 1-ethyl-3-methylimidazolium tetrafluoroborate (EMI-BF$_4$). Meniscus size and operational voltage were kept the same, while the hydraulic impedance factor is considered to be inversely proportional to viscosity $\mu$.

\begin{figure}
    \centering
    \includegraphics[width=1.02\linewidth]{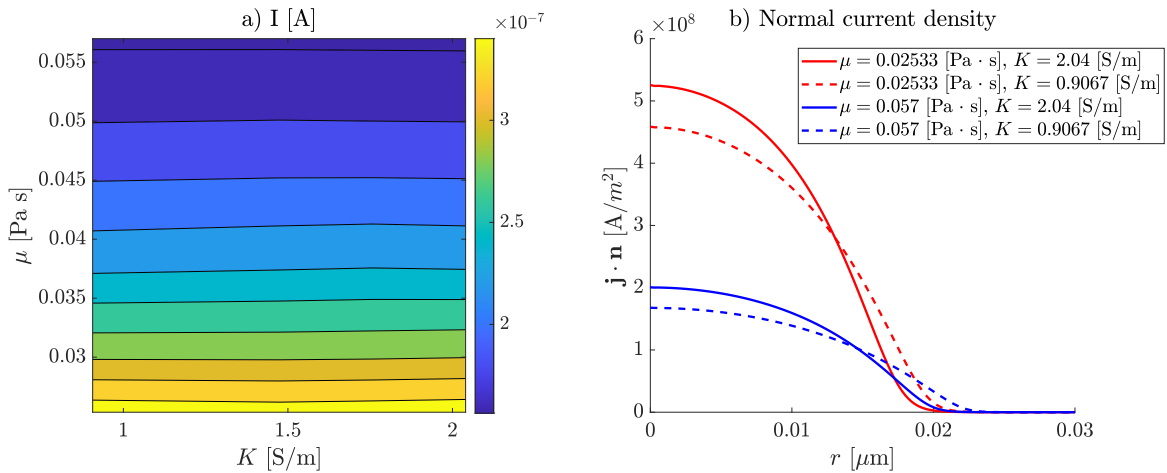}\caption{(a) Representative map showing the emitted current in a parameter space of varying  conductivities $K$ and viscosities of the liquid, and (b) axisymmetric current density distribution along the emission region of the meniscus interface. The radius $r = 0$ corresponds to the apex of the meniscus or axis of symmetry. Around $r = 3 \cdot 10^{-8}$ m, the emission region ends for these simulated menisci.  Solid and dashed curves share the same conductivity at $K = 2.04$ S/m and $K = 0.91$ S/m, respectively. Red and blue curves share the same viscosity at $\mu = 0.025$ Pa$\cdot$s and $\mu = 0.057$ Pa$\cdot$s, respectively.}
    \label{fig:viscosityconductivity}
\end{figure}

The main observation is that the current emitted from the ionic liquid surface has practically no dependence on conductivity, as seen from the nearly-flat iso-current lines shown in figure \ref{fig:viscosityconductivity}a. The scaling in eqs.\ref{eq:rstar_2} and \ref{eq:current_density} might be true very locally, but will adjust slightly differently over different portions of the emission region. This is actually observed in the computed distributions of properties over the emission region in an electrically stressed meniscus.

\begin{figure}
    \centering
    \includegraphics[width=0.8\textwidth]{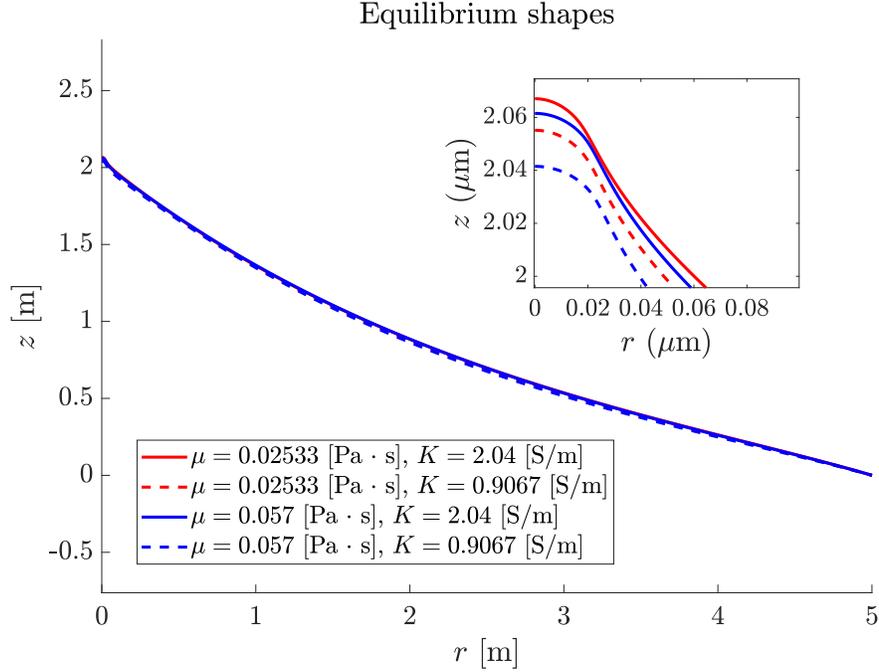}
    \caption{Equilibrium shapes for the limit cases presented in figure \ref{fig:viscosityconductivity}b.}
    \label{fig:eqshapes}
\end{figure}

As a pertinent example, the current densities for different viscosities are shown in figure \ref{fig:viscosityconductivity}b. The electrical conductivity has a clear effect on the shape of the current density $j=j(r,z)$ distributed over the meniscus surface. Even the shape of the meniscus is different very close to the emission region (figure \ref{fig:eqshapes}).
Higher conductivity leads to a higher current density near the apex, but at the same time, the active area from which ions are emitted becomes smaller.   Interestingly, these changes cancel out precisely, such that integrated quantities, in particular, the total emitted current,

$$
I = \int\limits_{\text{meniscus}} j\;dA,
$$

is independent of the value of electric conductivity $K$. As a consequence, the flow rate is also independent of electric conductivity, as in the cone-jet regime. The reduction of $r^*$ at higher conductivities in effect becomes similar to what is observed in cone-jet electrosprays with prescribed flow (eq. \ref{eq:jet_diameter}, if we assume that the diameter of the droplet is a lower bound for the diameter of the jet), except that now it is apparent that this result applies to situations where the flow rate is part of the solution. As such, evidence suggests that what determines the flow rate in passive sources are the electric field and the hydraulic impedance. 

The hydraulic impedance quantifies how much pressure drop per unit flow rate is originated from the propellant reservoir to the emitting meniscus due to fluid friction with the emitter walls. Hydraulic impedance is generally independent from conductivity, and affected by the viscosity of the liquid $\mu$, and the geometry of the emitter transporting the flow. This is supported by experimental evidence, for example, the work by Castro \cite{Castro2009EffectTips} and Krpoun \cite{Krpoun2009TailoringElectrodes}, where different ion sources wetted with the same liquid yielded much more current with tip geometries favoring lower impedance. The model predicts that this current increase is inversely proportional to the reduction in hydraulic impedance  \cite{Gallud2022TheEvaporation,Coffman2016}. 


Other examples come from the operational principles of liquid metal ion sources, which should be very similar to those of ionic liquid electrosprays in the ionic regime (ionic liquid ion sources). The surface stresses and flow conditions are similar, although their physical and electric properties are very different: liquid metals have, on average an order of magnitude higher surface tension and, remarkably, at least six orders of magnitude larger electric conductivity. If the scaling in eqs. \ref{eq:current_density}-\ref{eq:current_2} were somewhat accurate and applicable for both ionic liquids and liquid metals, then we should expect the emission current from liquid metals to be up to six orders of magnitude higher than what is obtained with ionic liquids. In reality, the onset currents achievable with liquid metals are up to $\sim 10^2$ times higher, a range that correlates much better with their difference in viscosities, and characteristic $\rho \dfrac{q}{m}$ values.

In fact, scaling laws derived for the current emitted in liquid metal ion sources do not depend on conductivity and charge relaxation time scales, nor any parameter regarding charge emission (e.g, $E^*$), or referencing current density. Quoting directly from Forbes \cite{Forbes2017LiquidSources}:
\begin{quote}
    ... steady-state LMIS i–V behavior can be explained by a theory based on total-force
arguments, the slender-body approximation, and the presence of space-charge, without any need
to resort to detailed arguments concerning apex field and current density\footnote{Forbes refers here to the field at the apex of the liquid metal meniscus, e.g, emission region $\approx E^*$}. Although this may seem
counterintuitive, it is characteristic of space-charge-controlled situations that emission details are
of limited importance.
\end{quote}

Results from our electrohydrodynamic modeling seem to show that such condition holds also when space charge effects do not dominate. 

\section{Experimental evidence}
In other works, where sources of similar emitter geometry were wetted with liquids of different conductivity \cite{Garoz2007TaylorConductivity}, the specific role of viscosity over conductivity could easily have gone unnoticed when trying to explain the enhanced currents observed in liquids of higher conductivity. The main reason is that conductivity and viscosity in ionic liquids are roughly correlated as their product $\mu K$ is approximately constant (Walden rule \cite{Schreiner2010FractionalPlot}), therefore higher conductivity ionic liquids tend to have also low viscosity. 
With this in mind, we provide experimental evidence that circumvents this correlation.


We have prepared two ionic liquid porous carbon xerogel sources of similar pore size, porosity, and height that produce ion beams from two different ionic liquids [EMIm][HSO$_4$] and [bis(mim)C$_6$][NTF$_2$]$_2$. The two liquids share similar viscosities at the temperature tested, but with more than two orders of magnitude difference in conductivity (see table \ref{tab:table}). The pure ionic current emitted by the lower conductivity liquid [EMIm][HSO$_4$] is substantially higher than for the higher conductivity ionic liquid (figure \ref{fig:comparison}a), and the ratio of both currents appears to correlate much better with the characteristic $\rho q/m$ values as seen from the flow rate plot in figure \ref{fig:comparison}b.
\section{The role of conductivity: enhancing the stability of the pure-ion regime}
The excitement has to be tempered however, since the situation is not as simple. While the evidence presented here suggests that the electric conductivity does not affect the magnitude of the emitted current, the same model predicts that lower conductivities reduce the range of flow rates in which menisci in the pure ionic regime are statically stable, e.g., when the liquid-vacuum interface is in perfect balance between the electric, surface tension and hydraulic stresses. This would explain why high impedance sources are especially needed in ionic liquids with low conductivity: to help decrease the flow rate within the ranges where the pure ionic current is sustainable at these conductivities \cite{Castro2006CapillarySources}, \cite{Fedkiw2009DevelopmentApplications}. It would also explain why a higher fraction of ion vs. droplet current is observed when the conductivity is increased by heating the ionic liquid in the mixed regime at constant flow rate \cite{Gamero-Castano2021ElectrospraysSpectrometry}.

\section{Conclusion}

Theoretical and experimental evidence suggest
that electric conductivity does not control the magnitude of the current emitted from ionic liquid ion sources. At the end, the property that enhances current in ionic liquid electrosprays seems to be low viscosity, with electric conductivity not being important. There have been clear hints that support this view that in retrospect now seem quite clear, but have passed nevertheless unnoticed: the current in cone-jet electrosprays is related to conductivity only because the specific charge depends on it. In the pure-ion mode, however, the specific charge has a well-defined value due to the discrete nature of emitted ions, therefore independent of the electric conductivity. On the other hand, increasing the current with low conductivity fluids requires attention to other liquid properties and operational conditions. 

Firstly, higher emitted currents could be achieved by selecting substances with high mass density and charge-to-mass ratios, and, perhaps more relevant that anything else, by increasing the flow rate through a reduction of viscosity (since $I = \rho Q\dfrac{q}{m}$). The differences in current magnitudes between LMIS and ILIS have much better correlation to viscosity $Q \propto \frac{1}{\mu}$ than scalings including conductivity ratios between the two liquids.

Secondly, strategies to expand the region of stability of electrically-stressed menisci would be required to improve the current emission capability, such as increasing the surface tension (which is what happens with liquid metals) or through precise manufacturing or selection of anchoring sites on the emitter structures that promote the formation of a meniscus size that maximizes the range of stability for a given hydraulic impedance.

Increasing the conductivity $K$ has been always a hard requirement to meet as only a relatively small number of liquids are available with $K>1$ S/m. On the other hand, there are thousands, perhaps millions, of liquids and certainly an infinity of mixtures, that could be used and are characterized by having lower $K$. The lack of a strong influence of electric conductivity on the magnitude of the emission current from ionic liquids opens up new lines of research and application opportunities. 
\begin{figure}
    \centering
    \includegraphics[width=0.8\linewidth]{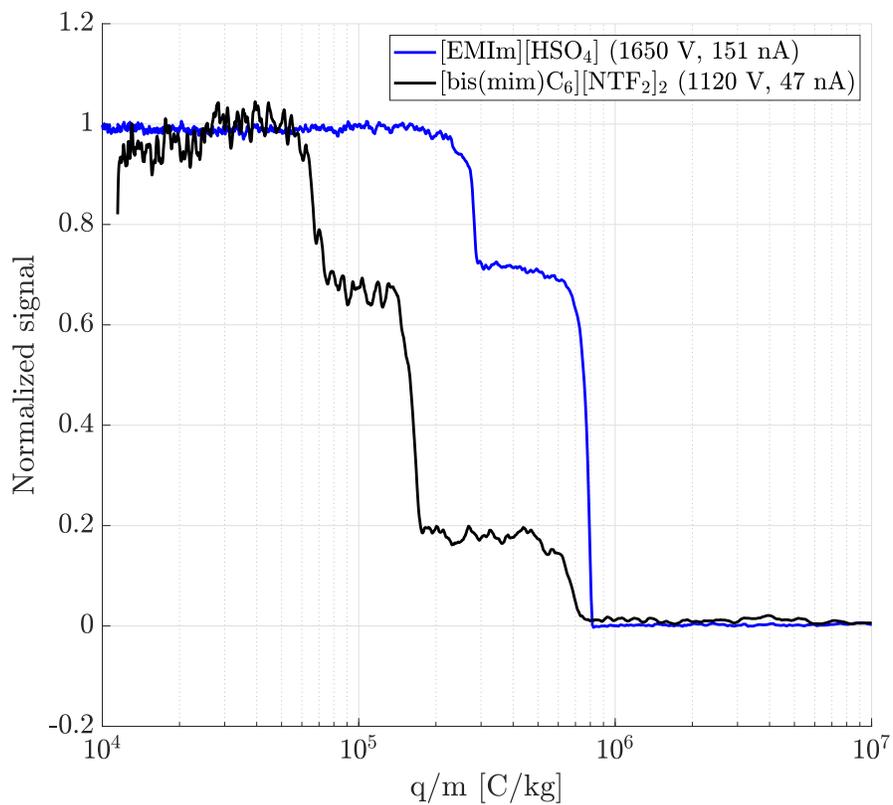}
    \caption{Time-of-flight polydispersity characterization of [bis(mim)C$_6$][NTF$_2$]$_2$ and [EMIm][HSO$_4$] at the voltages tested.}
    \label{fig:tof}
\end{figure}
\begin{landscape}
\begin{figure}
    \centering
    \includegraphics[width=0.8\linewidth]{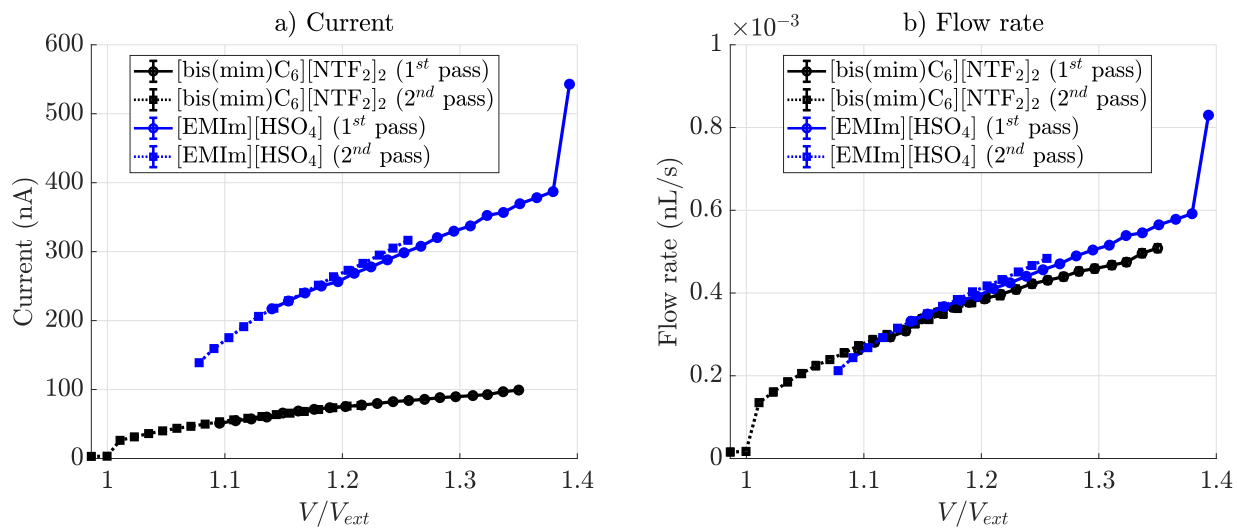}
    \caption{Current and flow rate vs. voltage curves of [bis(mim)C$_6$][NTF$_2$]$_2$ and [EMIm][HSO$_4$]. Curves scaled by the extinction voltage $V_{ext}$.}
    \label{fig:comparison}
\end{figure}
\begin{table}
\begin{tabular}{lcccccc}%
\toprule
\multirow{2}{*}{Ionic liquid} & 
Temperature
& 
Viscosity  &
Electrical& 
Surface &
Density &
Average $q/m$ 
\tablefootnote{From time of flight spectroscopy (see figure \ref{fig:tof})}\\
& ($^{\circ}$C)
& (Pa s)
& cond. (S/m)
& tension (N/m)
& (kg/m$^3$)
& (C/kg) \\
\midrule
$[$EMIm$]$$[$HSO$_4]$ & 47.8 & 0.355 \cite{Garcia-Garabal2017TRANSPORTMECHANISM}& 0.305 \cite{Garcia-Garabal2017TRANSPORTMECHANISM} & 0.057 \cite{Jianghua2007SynthesisAluminum} & 1360 \cite{Garcia-Garabal2017TRANSPORTMECHANISM}&    4.81$\cdot 10^5$
   \\
$[$bis(mim)C$_6][$NTF$_2$]$_2$ & 30 & 0.357 \cite{Zhang2013AbsorptionMechanisms} & 107.4 \cite{Zhang2013AbsorptionMechanisms} & 0.04 \tablefootnote{In-house measurement} & 1525 \cite{Zhang2013AbsorptionMechanisms} & 0.9$\cdot10^5$\\
\midrule
\bottomrule
\end{tabular}
\caption{\label{tab:table}Properties and testing conditions of [bis(mim)C$_6$][NTF$_2$]$_2$ and [EMIm][HSO$_4$].}
\end{table}
\end{landscape}

Funding for this research was provided by the Air Force Office of Scientific Research through research grant FA9550-19-1-0104 monitored by Dr. Mitat Birkan.

\bibliography{references}

\begin{thebibliography}{10}

\bibitem{Iribarne1976}
J.~V. Iribarne and B.~A. Thomson.
\newblock {On the evaporation of small ions from charged droplets}.
\newblock {\em The Journal of Chemical Physics}, 64(6):2287, 1976.

\bibitem{Romero-Sanz2003SourceRegime}
I.~Romero-Sanz, R.~Bocanegra, Juan Fern{\'{a}}ndez de~la Mora, and Manuel
  Gamero-Casta{\~{n}}o.
\newblock {Source of heavy molecular ions based on Taylor cones of ionic
  liquids operating in the pure ion evaporation regime}.
\newblock {\em Journal of Applied Physics}, 94(5):3599--3605, 2003.

\bibitem{Romero-Sanz2005IonicLiquids}
I.~Romero-Sanz, I.~Aguirre~de C{\'{a}}rcer, and Juan Fern{\'{a}}ndez de~la
  Mora.
\newblock {Ionic propulsion based on heated Taylor cones of ionic liquids}.
\newblock {\em Journal of Propulsion and Power}, 21(2):239--242, 2005.

\bibitem{Garoz2007TaylorConductivity}
D.~Garoz, C.~Bueno, C.~Larriba, S.~Castro, I.~Romero-Sanz, J.~Fernandez
  De~La~Mora, Y.~Yoshida, and G.~Saito.
\newblock {Taylor cones of ionic liquids from capillary tubes as sources of
  pure ions: The role of surface tension and electrical conductivity}.
\newblock {\em Journal of Applied Physics}, 102(6):064913, 9 2007.

\bibitem{Lozano2006EnergySource}
Paulo~C. Lozano.
\newblock {Energy properties of an EMI-Im ionic liquid ion source}.
\newblock {\em Journal of Physics D: Applied Physics}, 39(1):126--134, 2006.

\bibitem{Castro2007EffectSourcesb}
S.~Castro, C.~Larriba, J.~Fernandez De~La~Mora, Paulo~C. Lozano, S~S{\"{u}}mer,
  Y~Yoshida, and G~Saito.
\newblock {Effect of liquid properties on electrosprays from externally wetted
  ionic liquid ion sources}.
\newblock {\em Journal of Applied Physics}, 102(9):44104, 2007.

\bibitem{Larriba2007MonoenergeticLiquids}
C.~Larriba, S.~Castro, J.~Fernandez De~La~Mora, and Paulo~C. Lozano.
\newblock {Monoenergetic source of kilodalton ions from Taylor cones of ionic
  liquids}.
\newblock {\em Journal of Applied Physics}, 101(8):084303, 4 2007.

\bibitem{Castro2006CapillarySources}
S~Castro, C.~Larriba, Juan Fern{\'{a}}ndez de~la Mora, Paulo~C. Lozano, and
  S~Sumer.
\newblock {Capillary vs. externally wetted ionic liquid ion sources}.
\newblock In {\em Collection of Technical Papers - AIAA/ASME/SAE/ASEE 42nd
  Joint Propulsion Conference}, volume~4, pages 3262--3267, 2006.

\bibitem{Chiu2007IonicPropulsion}
Yu~Hui Chiu and Rainer~A. Dressler.
\newblock {Ionic liquids for space propulsion}.
\newblock In {\em ACS Symposium Series}, volume 975, pages 138--160, 2007.

\bibitem{Takeuchi2013DevelopmentModification}
Mitsuaki Takeuchi, Takuya Hamaguchi, Hiromichi Ryuto, and Gikan~H. Takaoka.
\newblock {Development of ionic liquid ion source with porous emitter for
  surface modification}.
\newblock {\em Nuclear Instruments and Methods in Physics Research, Section B:
  Beam Interactions with Materials and Atoms}, 315:345--349, 11 2013.

\bibitem{Guilet2011IonicMachining}
S.~Guilet, C.~Perez-Martinez, P.~Jegou, P.~Lozano, and J.~Gierak.
\newblock {Ionic liquid ion sources for silicon reactive machining}.
\newblock {\em Microelectronic Engineering}, 88(8):1968--1971, 2011.

\bibitem{Yoshida2014ConcurrentArray}
Ryo Yoshida, Motoaki Hara, Hiroyuki Oguchi, Tatsuya Suzuki, and Hiroki Kuwano.
\newblock {Concurrent reactive ion etching employing micromachined ionic liquid
  ion source array}.
\newblock In {\em Proceedings of the IEEE International Conference on Micro
  Electro Mechanical Systems (MEMS)}, pages 463--466. Institute of Electrical
  and Electronics Engineers Inc., 2014.

\bibitem{Perez-Martinez2015IonMicrotips}
Carla P{\'{e}}rez-Mart{\'{i}}nez and Paulo~C. Lozano.
\newblock {Ion field-evaporation from ionic liquids infusing carbon xerogel
  microtips}.
\newblock {\em Applied Physics Letters}, 107(4), 7 2015.

\bibitem{Castro2009EffectTips}
S~Castro and Juan Fern{\'{a}}ndez de~la Mora.
\newblock {Effect of tip curvature on ionic emissions from Taylor cones of
  ionic liquids from externally wetted tungsten tips}.
\newblock {\em Journal of Applied Physics}, 105(3), 2009.

\bibitem{Smith2006VoltageElectrospraying}
K.~L. Smith, M.~S. Alexander, and J.~P.W. Stark.
\newblock {Voltage effects on the volumetric flow rate in cone-jet mode
  electrospraying}.
\newblock {\em Journal of Applied Physics}, 99(6):064909, 3 2006.

\bibitem{Smith2006TheIN}
Katharine~L Smith, Matthew~S Alexander, and John P~W Stark.
\newblock {The sensitivity of volumetric flow rate to applied voltage in
  cone-jet mode electrospray and the influence of solution properties and
  emitter geometry ARTICLES YOU MAY BE INTERESTED IN}.
\newblock {\em Phys. Fluids}, 18:92104, 2006.

\bibitem{Ryan2014TheModes}
C~N Ryan, K~L Smith, and J~P~W Stark.
\newblock {The flow rate sensitivity to voltage across four electrospray
  modes}.
\newblock {\em Citation: Applied Physics Letters}, 104:44905, 2014.

\bibitem{Ganan-Calvo1999TheLawsb}
Alfonso~M. Ga{\~{n}}{\'{a}}n-Calvo.
\newblock {The surface charge in electrospraying: Its nature and its universal
  scaling laws}.
\newblock {\em Journal of Aerosol Science}, 30(7):863--872, 8 1999.

\bibitem{FernandezdelaMora1994TheCones}
Juan Fern{\'{a}}ndez de~la Mora and Ignacio~G. Loscertales.
\newblock {The current emitted by highly conducting Taylor cones}.
\newblock {\em Journal of Fluid Mechanics}, 260(special issue):155--184, 1994.

\bibitem{Ganan-Calvo1997CurrentLaws}
Alfonso~M. Ga{\~{n}}{\'{a}}n-Calvo, Jonathan~M. D{\'{a}}vila, and Antonio
  Barrero.
\newblock {Current and droplet size in the electrospraying of liquids. Scaling
  laws}.
\newblock {\em Journal of Aerosol Science}, 28(2):249--275, 1997.

\bibitem{Higuera}
Francisco~J. Higuera.
\newblock {Model of the meniscus of an ionic-liquid ion source}.
\newblock {\em Physical Review E - Statistical, Nonlinear, and Soft Matter
  Physics}, 77(2), 2008.

\bibitem{Coffman2019ElectrohydrodynamicsField}
Chase~S. Coffman, Manuel Mart{\'{i}}nez-S{\'{a}}nchez, and Paulo~C. Lozano.
\newblock {Electrohydrodynamics of an ionic liquid meniscus during evaporation
  of ions in a regime of high electric field}.
\newblock {\em Physical Review E}, 99(6):063108, 2019.

\bibitem{Gallud2022TheEvaporation}
Ximo Gallud and Paulo~C. Lozano.
\newblock {The emission properties, structure and stability of ionic liquid
  menisci undergoing electrically assisted ion evaporation}.
\newblock {\em Journal of Fluid Mechanics}, 933, 2 2022.

\bibitem{Krpoun2009TailoringElectrodes}
R.~Krpoun, K.~L. Smith, J.~P.W. Stark, and Herbert~R. Shea.
\newblock {Tailoring the hydraulic impedance of out-of-plane micromachined
  electrospray sources with integrated electrodes}.
\newblock {\em Applied Physics Letters}, 94(16):163502, 4 2009.

\bibitem{Coffman2016}
Chase~S. Coffman.
\newblock {\em {Electrically-Assisted Evaporation of Charged Fluids:
  Fundamental Modeling and Studies on Ionic Liquids}}.
\newblock PhD thesis, Massachusetts Institute of Technology, 2016.

\bibitem{Forbes2017LiquidSources}
Richard~G. Forbes and Graeme~L.R. Mair.
\newblock {Liquid metal ion sources}.
\newblock In {\em Handbook of Charged Particle Optics, Second Edition}, pages
  29--86. CRC Press, 12 2017.

\bibitem{Schreiner2010FractionalPlot}
Christian Schreiner, Sandra Zugmann, Robert Hartl, and Heiner~J. Gores.
\newblock {Fractional walden rule for ionic liquids: Examples from recent
  measurements and a critique of the so-called ideal KCl line for the walden
  plot}.
\newblock {\em Journal of Chemical and Engineering Data}, 55(5):1784--1788, 5
  2010.

\bibitem{Fedkiw2009DevelopmentApplications}
Timothy~P. Fedkiw and Paulo~C. Lozano.
\newblock {Development and characterization of an iodine field emission ion
  source for focused ion beam applications}.
\newblock {\em Journal of Vacuum Science {\&} Technology B: Microelectronics
  and Nanometer Structures}, 27(6):2648, 2009.

\bibitem{Gamero-Castano2021ElectrospraysSpectrometry}
Manuel Gamero-Casta{\~{n}}o and Albert Cisquella-Serra.
\newblock {Electrosprays of highly conducting liquids: A study of droplet and
  ion emission based on retarding potential and time-of-flight spectrometry}.
\newblock {\em Physical Review Fluids}, 6(1):013701, 1 2021.

\bibitem{Garcia-Garabal2017TRANSPORTMECHANISM}
S.~Garc{\'{i}}a-Garabal, J.~Vila, E.~Rilo, M.~Dom{\'{i}}nguez-P{\'{e}}rez,
  L.~Segade, E.~Tojo, P.~Verd{\'{i}}a, L.~M. Varela, and O.~Cabeza.
\newblock {TRANSPORT PROPERTIES FOR 1-ETHYL-3-METHYLIMIDAZOLIUM n-ALKYL
  SULFATES: POSSIBLE EVIDENCE OF GROTTHUSS MECHANISM}.
\newblock {\em Electrochimica Acta}, 231:94--102, 3 2017.

\bibitem{Jianghua2007SynthesisAluminum}
MA~Jianghua, LI~Yuping, LI~Huiquan, and ZHANG Yi.
\newblock {Synthesis of 1-Ethyl-3-methylimidazolium Hydrogen Sulfate and Its
  Application in the Electrolysis of Aluminum}.
\newblock {\em Chinese Journal of Process Engineering}, 7(6):1083--1088, 2007.

\bibitem{Zhang2013AbsorptionMechanisms}
Yi~Zhang, Ping Yu, and Yunbai Luo.
\newblock {Absorption of CO2 by amino acid-functionalized and traditional
  dicationic ionic liquids: Properties, Henry's law constants and mechanisms}.
\newblock {\em Chemical Engineering Journal}, 214:355--363, 1 2013.

\end{thebibliography}
\bibliographystyle{unsrt}

\end{document}